\documentclass[preprint,12pt]{aastex}
\usepackage{amsmath}
\usepackage{graphicx}
\usepackage{natbib}

\newcommand{\gm}{\gamma_{max}}
\newcommand{\g}{\gamma}
\newcommand{\G}{\Gamma}
\newcommand{\epsBd}{\varepsilon_{B,d,-2}}
\newcommand{\epse}{\varepsilon_e}

\begin{document}

\title{On particle acceleration rate in GRB afterglows}
\author{Eran Sagi and Ehud Nakar }
\affil{Raymond and Beverly Sackler School of Physics \& Astronomy,
Tel Aviv University, Tel Aviv 69978, Israel\\}

\begin{abstract}
It is well known that collisionless shocks are major sites of
particle acceleration in the Universe, but the details of the
acceleration process are still not well understood. The particle
acceleration rate, which can shed light on the acceleration process,
is rarely measured in astrophysical environments. Here we use
observations of gamma-ray burst afterglows, which are weakly
magnetized relativistic collisionless shocks in ion-electron plasma,
to constrain the rate of particle acceleration in such shocks. We
find, based on X-ray and GeV afterglows,  an acceleration rate that
is most likely very fast, approaching the Bohm limit, when the shock
Lorentz factor is in the range of $\G \sim 10-100$. In that case
X-ray observations may be consistent with no amplification of the
magnetic field in the shock upstream region. We examine the X-ray
afterglow of GRB 060729, which is observed for 642 days showing a
sharp decay in the flux starting about 400 days after the burst,
when the shock Lorentz factor is $\sim 5$. We find that inability to
accelerate X-ray emitting electrons at late time provides a natural
explanation for the sharp decay, and that also in that case
acceleration must be rather fast, and cannot be more than a 100
times slower than the Bohm limit. We conclude that particle
acceleration is most likely fast in GRB afterglows, at least as long
as the blast wave is ultra-relativistic.

\end{abstract}

\section{Introduction}
Astrophysical collisionless shocks are efficient particle
accelerators. The signature of ultra-relativistic particles that are
accelerated in these shocks is seen in a variety of astrophysical
phenomena and over a wide range of environments. Nevertheless,
despite of an extensive study, the acceleration processes are still
largely unknown. One of the leading candidates is the diffusive
shock acceleration (DSA; e.g.,
\citealt{Bell78,Blandford1978,Blandford87}), where charged particles
are accelerated by crossing the shock back and forth. The
acceleration time in DSA depends on the duration that it takes a
particle to close a single cycle, i.e., to cross the shock back and
forth one time. This time increases with the particle energy, and it
sets the maximal Lorentz factor, $\gm$ that a particle can achieve.
Thus, measuring $\gm$ provides a direct information about the
acceleration process and about the physical conditions in the
acceleration site.

The reflection of particles back and forth through the shock is
believed to be done by scattering on fluctuating magnetic fields.
This process is typically approximated as a diffusion in direction
of the particles velocity, in which case the duration of the
acceleration depends on the particle mean free path, $\lambda$. It
is reasonable to assume that typically the shortest possible mean
free path is the Larmor radius $r_l$ (Bohm limit), and thus to
parameterize the diffusion by $\lambda=\eta r_l$ where $\eta \geq 1$
is generally expected, although this is not a hard lower limit. The
value of $\eta$, which measures the diffusion efficiency and how
fast is the acceleration, was constrained only in a small number of
systems. Probably the best estimate is obtained from the gamma-ray
spectrum of the Crab nebula \citep{deJager92}. The spectrum shows
two components, where the lower energy component, which is most
likely dominated by synchrotron, shows a cut-off around 100 MeV. The
fact that synchrotron emission reaches these energies, given the
rapid synchrotron cooling, implies $\eta \approx 1$. Higher energy
synchrotron emission during flares suggest that maybe even $\eta <
1$ is required \citep{Abdo11}. Another system where $\eta$ was
claimed to be measured is the supernova remnant SNR RXJ1713.72-3946,
where \cite{Uchiyama07} find that observed X-ray variability on a
year time scale indicates on $\eta \sim 1$. Thus, two very different
acceleration sites, one relativistic, possibly highly magnetized
shock in pair plasma and the other Newtonian shock in ion-electron
plasma suggest that particle acceleration, if dominated by DSA, is
extremely fast.  Calculating $\eta$ from first principles is
impossible at this point, since it depends on the unknown shock
structure, and most importantly on the properties of the upstream
and downstream magnetic fields. Calculations of $\eta$ in
relativistic shocks were done only by assuming the magnetic field
structure. For example, \cite{Lemoine03} and \cite{Lemoine06} find
that when a Kolmogorov magnetic turbulence spectrum is assumed in
the upstream region, then $\eta \approx 10$.

Here we examine the constraints that can be obtained on  $\eta$ from
observations of cut-off, or the lack thereof, in GRB afterglow light
curves and spectra. These afterglows are almost certainly generated
by a relativistic blast waves that propagate into a weakly
magnetized ion-electron plasma (for reviews see
\citealt{Piran04,Meszaros06,Nakar07}). Previous studies of particle
acceleration in GRB afterglows assumed $\eta=1-10$ and used the lack
of spectral cut-off in observed X-ray and GeV emission to constrain
the magnetic field upstream and/or downstream of the shock
(\citealt{Li06}; \citealt{Piran10} [PN10];
\citealt{BarniolDuran11a,Li11}). Here we take a different approach
asking how well $\eta$ can be constraint. Moreover, we find that the
fast decay observed in the extraordinarily year long X-ray afterglow
of GRB 060729 is naturally explained by the inability of the shock
to accelerate X-ray emitting electrons, providing a measurement of
$\eta$ in that case.

In section \S\ref{sec gamma limits} we describe the various limits
on the $\gm$. The resulting limits on $\eta$, for various circum
burst density profiles, are derived in \S\ref{sec eta limits}. The
special case of the GRB with the longest duration X-ray afterglow,
GRB 060729, is discussed in \S\ref{sec GRB 060729}.

\section{Limits on the maximal Lorentz factor}\label{sec gamma limits}
The two main factors that limit the acceleration of particles in a
decelerating relativistic blast wave are confinement and cooling
\citep[e.g.,][]{Li06,Piran10,BarniolDuran11a}. Below we shortly
discuss these limits (see PN for detail). Observations indicate that
the magnetic field in the downstream region is amplified in GRB
external shocks well beyond the effect of compression. Thus, most of
the emission take place in the downstream region while a particle
spends most of its acceleration time in the upstream region.
Throughout the paper we assume that $\eta$ is similar in the shock
downstream and upstream, but we highlight which of the observations
constrain $\eta$ in downstream region and which constrain $\eta$ in
the upstream region.

Confinement is limited by the ability of the accelerated particle
that is moving in the upstream region to cross the shock back into
the downstream region. Thus, confinement is limited by $\eta$ in the
upstream region. Its limit on the maximal Lorentz factor is set by
the requirement that the particle complete a turn of 180$^o$, as
seen in the shock frame, while the shock propagates a distance $f_u
R$, where $f_u$ accounts for the shock deceleration (see PN10).
Thus,
\begin{equation}\label{eq: gamma_conf}
    \g'_{conf} \approx \frac{e B_u}{\eta m_e c^2} f_u R
\end{equation}
where `` $'$ " denotes quantities in the shock rest frame, $R$ is
the shock radius, $e$ and $m_e$  are the electron charge and mass, c
is the speed of light and $B_u$ is the rest frame upstream magnetic
field.

The two processes that dominate cooling are synchrotron and inverse
Compton (IC). In both cases the maximal Lorentz factor is found by
equating the acceleration time (i.e., the time to complete a Fermi
cycle in relativistic shocks) to the relevant cooling time. In case
of synchrotron:
\begin{equation}\label{eq: gamma_synch}
    \g'_{synch} \approx \left(\frac{6 \pi e}{\eta B_d
    \sigma_T}\right)^\frac{1}{2} ,
\end{equation}
where $\sigma_T$ is the Thomson cross-section and $B_d$ is the
downstream magnetic field. Since observations indicate that the
downstream magnetic field is amplified by the shock, synchrotron
cooling in the downstream region is more limiting than in the
upstream region and it sets a limit on $\eta$ in the downstream
region.

Inverse Compton cooling is more efficient in the upstream region,
since the radiation field is similar in both sides of shock, but a
particle spends more time in the upstream region, where the magnetic
field is lower. Thus, $\g'_{IC}$ is limited by $\eta$ in the
upstream region:
\begin{equation}\label{eq: gamma_IC}
    \g'_{IC} \approx \left(\frac{3 e B'_u}{4\eta
     \sigma_T U'_{rad}(<\nu'_{KN})}\right)^\frac{1}{2}= \g'_{synch} \left( \frac{B'_u}{Y(\g'_{IC}) B_d }\right)^{1/2} ,
\end{equation}
where $B'_u$ is the magnetic field in the upstream region as
measured in the shock frame (related to the rest frame magnetic
field by $B'_u \approx \G B_u$). $U'_{rad}(<\nu'_{KN})$ is the shock
frame radiation energy density at frequencies smaller than:
\begin{equation}\label{eq nu_kn}
    \nu'_{KN}(\g'_{IC})=\frac{m_e c^2}{h \g'_{IC}} ,
\end{equation}
where $h$ is the Planck constant. Equation \ref{eq: gamma_IC} gives
also the relation between $\g'_{IC}$ and $\g'_{synch}$  using the
ratio between IC and synchrotron cooling rate in the downstream
region, $Y(\g'_{IC}) = U'_{rad}(<\nu'_{KN})/(B_d^2/8\pi)$.

\section{Maximal observed frequency and limits on $\eta$}\label{sec eta limits}
Below we derive the constraints that X-ray and GeV afterglows set on
$\eta$. We assume that the observed emission is synchrotron
radiation generated by a quasi-spherical decelerating adiabatic
blast wave. This is almost certainly the case in many X-ray
afterglows, at least during the first day. The origin of the
observed long lasting GeV emission, which is seen up to $\sim 1000$
s after some bursts, is still unclear, although observations suggest
that it is also synchrotron emission from the decelerating blast
wave \citep{Kumar09,Kumar10,Ghisellini10}. Since afterglow
observations suggest that the circum-burst density profile vary from
one GRB to another, we consider here two typical external density
profiles, one constant, as expected for the interstellar medium
(ISM) and one $\propto R^{-2}$ as expected for a stellar wind.

\subsection{ISM}
Under the assumption of spherical expansion in a constant density
$n$, the radius and Lorentz factor of an adiabatic blast wave with
energy $E$ at an observer time $t$ are
\citep[e.g.,][]{Saripirannarayan98}:
\begin{equation}
\begin{array}{c}
  R \approx 6 \cdot 10^{17}{\rm~cm} ~\left(\frac{E_{53}}{n}\right)^{1/4}\left(\frac{1+z}{3}\right)^{-1/4} t_5^{1/4} \\
  \\
  \G  \approx 12 \left(\frac{E_{53}}{n}\right)^{1/8}\left(\frac{1+z}{3}\right)^{3/8} t_5^{-3/8}
\end{array}
\end{equation}
where $z$ is the burst redshift and $q_x$ denotes the value of
$q/10^x$ in c.g.s. units.  We assume that the downstream magnetic
field is a constant fraction, $\varepsilon_{B,d}$, of the internal
energy behind the shock so $B_{d} \approx \left( 32\pi
\varepsilon_{B,d} n \G^2 m_p c^2 \right)^{1/2}$, where $m_p$ is the
proton mass. If the magnetic field in the upstream is not amplified
by a precursor to the shock then it is expected to be constant and
of order of $10\mu G$. If it is amplified then it may be
significantly larger. Finally, $f_u = 1/3$ in ISM (PN10). Using the
synchrotron emission from the downstream region, $h \nu = \G \g'^2 e
B_d /(2\pi m_e c)$, we obtain the maximal frequency that is dictated
by the limits discussed above:
\begin{align}\label{eq hnu_ISM}
&h{{\nu }_{conf}} \approx 2 \cdot 10^{10} {\rm ~eV} ~ \eta^{-2}
E_{53}^{3/4}\epsBd^{1/2}B_{u,-5}^{2}n_{0}^{-1/4}t_{5}^{-1/4} \nonumber \\
&h{{\nu }_{sync}}\approx 2.5 \cdot {{10}^{8}} {\rm ~eV} ~ \eta^{-1}
t_{5}^{-3/8} \\
&h{{\nu }_{IC}}\approx 1.5 \cdot 10^{5} {\rm ~eV} ~ \eta^{-1} B_{u,-5} \epsBd^{-1/2}n_0^{-5/8}{{Y}^{-1}}t_{5}^{-3/8} \nonumber
\end{align}
where here, and throughout the paper, we derive values for the
typical $z=2$. We also ignore dependencies on parameters that are
raised to the power of $1/8$ since these cannot affect the result by
an order of magnitude, which is the accuracy of our calculation to
begin with. The constraints that we derive for confinement and
synchrotron are similar to those of PN10, which assumed $\eta=1$,
and the constraint on the IC cooling is similar to the one derived
in \cite{Li06}, which used a canonical value of $\eta=10$.

Equation \ref{eq hnu_ISM} implies that for canonical GRB parameters
confinement does not play an important role when the afterglow blast
wave propagates into ISM, with the possible exception of very early
time GeV emission\footnote{For early time GeV photons, the
confinement limit can be of the same order of magnitude as the
synchrotron cooling limit. However, it is less robust than
synchrotron limit due to poorly constrained parameters such as
$B_{u}$. Note, that in specific cases, confinement may provide the
most stringent constraint. An example is the extreme case where the
magnetic field is assumed not to be amplified by the shock and
$\varepsilon_{B,d} \sim 10^{-8}$ (\citealt{Kumar09}, PN10.)}. The
limit provided by synchrotron cooling is the most robust as it is
independent of almost anything\footnote{The synchrotron limit on the
rest frame frequency is independent of anything and is in general $h
\nu' \approx 50$MeV \citep{deJager92,Lyutikov09,Kirk10}. The
observed frame limit depends only on the Lorentz boost and
cosmological redshift, which vary by less than an order of magnitude
for reasonable values of $E$, $n$ and $z$. A frequency that is
higher by some factor may be obtained in configurations where the
acceleration takes place in a relatively weak magnetic field and
then the radiation takes place in a stronger field
\citep{Lyutikov09}. In that case the synchrotron limit on $\eta$ is
larger by that factor.}, except for the time since the explosion,
which is typically well measured:
\begin{equation}\label{eq eta_synchISM}
    \eta \lesssim 3 \left(\frac{h \nu_{obs} }{1 {~\rm Gev}}\right)^{-1} \left(\frac{t}{100 {~\rm
    s}}\right)^{-3/8} .
\end{equation}
Several Gev photons are seen $\sim 100$ s after the burst starts in
a number of the Fermi-LAT GRBs
\citep[e.g.,][]{Abdo09b,Abdo09a,Ackermann11} while $>100$ MeV
photons are seen in large numbers up to $\sim 1000$ s after the
burst in many Fermi-LAT GRBs. Therefore, if the Gev emission is
emitted by synchrotron process in the external shock, as suggested
by several authors \citep{Kumar09,Kumar10,Ghisellini10} then the
acceleration process in ultra-relativistic ($\G \sim 100$), weakly
magnetized, shock must be extremely fast\footnote{ Note that if the
blast wave is radiative, as suggested by \cite{Ghisellini10}, then
its Lorentz factor drops faster with time than in the adiabatic
case. Since the synchrotron limit on $\eta$ depends only on the
Lorentz factor, the limits that it provides in a case of a radiative
blast wave are tighter than those that we derive here for an
adiabatic blast wave.} with $\eta \lesssim 1$ in the shock
downstream.

The IC constraint depends on the value of the $Y$ parameter.
$Y_{GeV}$, the Y parameter of GeV emitting electrons, varies by many
orders of magnitude across the relevant phase space
\citep{Nakar09,Li11,BarniolDuran11b}. As a result the synchrotron
cooling limit is more stringent in part of the phase space.
Therefore, considering the robustness and tightness of the
synchrotron constraint, we do not attempt to cover here the possible
IC limits on the GeV emission. However, when considering X-ray
emission the synchrotron limit is very loose.  Therefore, we
consider the IC limit in that case, for which we need to evaluate
$Y_x$. This is not trivial due to Klein-Nishina [KN] effects that
play different roles over various areas of the phase space. An upper
limit on $Y_x$ can be easily obtained by assuming that the electrons
are in the fast cooling regime and that KN effects are negligible.
In that case $Y_x=\sqrt{\varepsilon_e/\varepsilon_{B,d}}$ if
$\varepsilon_e>\varepsilon_{B,d}$ \citep[e.g.,][]{Sari01}, where
$\varepsilon_e$ is the fraction of the internal energy behind the
shock that goes into accelerated electrons. For typical parameters
the electrons are cooling slowly at $t>10^4$ s, implying that $Y_x$
is smaller for electron distribution with a power-law index $p>2$:
$Y_x=\sqrt{\varepsilon_e/\varepsilon_{B,d}}
(\gamma_c/\gamma_m)^\frac{2-p}{2}$ were $\gamma_m$ is the typical
(also minimal) Lorentz factor of accelerated electrons and
$\gamma_c$ is the Lorentz factor of electrons that are cooling over
dynamical time scale. In addition, over a large range of the
parameter space KN suppression can be important, reducing the value
of $Y_x$ further. To account for these effects we use equations 46,
59, 60 and 63 of \cite{Nakar09}, which take consideration of the KN
effects and their feedback on the electron distribution, to scan the
phase space for the value of $\varepsilon_{B,d}^{1/2} Y_x$ (which
appears in the IC limit of equation \ref{eq hnu_ISM}). We scan the
parameter phase space and find that if the fraction of downstream
region internal energy that goes to electrons is $\epse=0.1$, the
electron distribution power-law index is in the range $p=2-2.8$
\citep{Curran10} and $\varepsilon_{B,d}>10^{-3}$, then the value of
$\epsBd^{1/2} Y_x$ is typically in the range of $0.3-3$ and its
dependence on the other parameters, $n$, $E$ and $t$, is rather weak
(most of the dependence in this range is on $p$ due to the fraction
of energy that is in fast cooling electrons, while KN suppression is
rather mild).
When $\varepsilon_B \ll 10^{-3}$ and/or $\epse \ll 0.1$
Klein-Nishina effects significantly suppress $Y_x$ and
$\varepsilon_B^{1/2} Y_x \ll 1$. We therefore conclude that for the
canonical values of $\varepsilon_B>10^{-3}$, $\epse=0.1$, $n_0
\approx 1$ and $p=2-2.8$
\begin{equation}\label{eq eta_IC_ISM}
    \eta \lesssim 15 ~\frac{10 {~\rm kev}}{h \nu_{obs}}B_{u,-5}
    t_{5}^{-3/8},
\end{equation}
 in the shock upstream region. If, however, $\varepsilon_B \ll
10^{-3}$ or $\epse\ll 0.1$ or $n_0 \ll 1$, then X-ray emission does
not provide strong constraints on $\eta$.

Many afterglows show X-ray emission (0.2-10 keV) that is bright for
days and in some cases weeks, without showing a clear sign of
spectral softening \citep{Liang08,Racusin09}. Thus, since afterglow
modeling typically implies $\varepsilon_B > 10^{-3}$ or $\epse
\approx 0.1$ and an ISM circum burst environment, observations of
X-ray afterglows suggest that the acceleration mancinism in
relativistic shocks ($\G \sim 10-50$) is fast. As evident from
equation \ref{eq eta_IC_ISM}, this limit depends on various
parameters. Some are constrained rather well, e.g., $\epse$, while
others are less constrained, e.g., $n_0$. Most important is the
dependence on $B_u$. There is a viable possibility that the
interaction of accelerated particles that run ahead of the shock
significantly amplifies the upstream magnetic field
\citep{Blandford87,Bell04,Milosavljevic06}. If this is the case then
the limits provided by X-ray observations are rather loose. In fact,
\cite{Li06} concluded, based on X-ray observations, that the
upstream magnetic field must be amplified at least up to $\gtrsim
0.1  n_0^{5/8}$mG. This conclusion was based on the assumption that
$\eta=10$.  They also take as a canonical value $\epsBd^{1/2} Y_x
\approx \sqrt{10}$, which they calculate by considering only the
part of the phase space where KN effects are negligible and by
taking $p=2$, for which, $Y_x$ is not suppressed by the slow cooling
of most of the
electrons. 
Our results show that a more careful estimate of $\epsBd^{1/2} Y_x$
reduces the \cite{Li06} limit by at least a factor of a few. In
addition, if the acceleration is as fast as suggested by the
recently detected GeV emission, and $\eta \sim 1$ also in the
upstream region then the limits on the upstream field drop to $\mu
G$ level, implying that current X-ray observations do not provide
strong evidence for magnetic field amplification in GRB afterglows.


\subsection {Wind}
The mass density profile in a wind from massive stars is $\rho=A
R^{-2}$. Under the assumption of spherical expansion the radius and
Lorentz factor of an adiabatic blast wave with energy $E$ at an
observer time $t$ are \citep{Chevalier00}:
\begin{equation}
\begin{array}{c}
  R=3 \cdot 10^{17}{\rm cm} \left(\frac{E_{53}}{A_{*}}\right)^{1/2}\left(\frac{1+z}{3}\right)^{-1/2}t_5^{1/2} \\
  \\
  \G =12 \left(\frac{E_{53}}{A_{*}}\right)^{1/4}\left(\frac{1+z}{3}\right)^{1/4}
  t_5^{-1/4}
\end{array}
\end{equation}
where $A_*=\frac{A}{5 \cdot 10^{11} {\rm~g/cm}}$ . Similarly to the
ISM case we assume that the downstream magnetic is a constant
fraction, $\epsBd$ of the internal energy behind the shock. Contrary
to the ISM case, the upstream magnetic field is not constant. The
magnetic field in the upstream region depends on the wind
magnetization and flux freezing implies $B_u \propto R^{-1}$,
assuming that upstream field is not amplified by the shock
precursor. The normalization depends on the wind velocity and on the
surface rotation velocity and magnetic field \citep{Goldreich70},
which are not tightly constrained. For typical parameters of a
Wolf-Rayet wind a field of $\sim 10\mu$G is expected at
$R=10^{18}$cm \citep{Eichler93}, but it can be more than an order of
magnitude larger or smaller. Therefore, we write the upstream
magnetic field as: ${{B}_{u}}=1\mu G{{\left( \frac{R}{{{10}^{19}}cm}
\right)}^{-1}}{{B}_{u,\mu G,19}}$. Using this parametrization  and
$f_u=1/2$ (PN10) the various constraints on the maximal observed
frequencies are:
\begin{align}\label{eq hnu_wind}
&h{{\nu }_{conf}}=3 \cdot {{10}^{11}} {\rm~eV} ~\eta^{-2} \epsBd^{1/2} A_*^{1/2} B_{u,\mu G,19}^2 t_5^{-1} \nonumber \\
&h\nu_{sync}=2 \cdot {{10}^{8}}{\rm~eV}~ \eta^{-1}
E_{53}^{1/4}A_{*}^{-1/4}t_{5}^{-1/4} \\
&h\nu_{IC}=3 \cdot {{10}^{5}}{\rm~eV}~ \eta^{-1}
E_{53}^{1/4}B_{u,\mu
G,19}\epsBd^{-1/2}A_{*}^{-3/4}{{Y}^{-1}}t_{5}^{-1/4} \nonumber
\end{align}

The most robust synchrotron limit is relevant only to the GeV
emission:
\begin{equation}\label{eq eta_synchWind}
    \eta \lesssim 1 \left(\frac{h \nu_{obs} }{1 {~\rm GeV}}\right)^{-1} \left(\frac{E_{53}}{A_*}\right)^{1/4}\left(\frac{t}{100 {~\rm
    s}}\right)^{-1/4} .
\end{equation}
Implying that if the observed GeV emission is produced by
synchrotron from an external shock in a wind environment then the
acceleration mechanism must be extremely fast. This limit is very
similar to the one obtained in the case of an ISM density profile
(equation \ref{eq eta_synchISM}), and is therefore general for any
reasonable circum burst density profile.

The IC limit depends on the value of the Y parameter. For the same
reasons discussed in the ISM case, we consider here IC limits only
on the X-ray emission. Unlike the ISM case the Y parameter of x-ray
emitting electrons, $Y_x$, depends strongly on time. In a wind
density profile the observed synchrotron cooing frequency, $\nu_c$
where most of the synchrotron energy is emitted, is increasing with
time. As a result, Klein-Nishina effects becomes significantly more
dominant with time, suppressing the IC cooling of X-ray emitting
electrons at late time.  The standard afterglow theory in a wind
\citep{Chevalier00} provides the value of $\nu_c(t)$  and of the
Lorentz factor of X-ray emitting electrons, $\g_x(t)$. Since at slow
cooling most of the synchrotron luminosity is emitted at $\nu_c$
(for $p<3$), KN effects are negligible for X-ray emitting electrons
as long as $\g_x \nu_c /\G < m_e c^2$. Thus, the time at which this
inequality becomes an equality provides a good approximation to the
time at which KN effects on $Y_x$ become important:
\begin{equation}\label{eq eta_synchWind}
    t_{Y_x,KN} \sim 4 \cdot 10^6 {\rm~s}~ A_*^{10/7} \epsBd^{4/7} \varepsilon_{e,-1}.
\end{equation}
This approximation assumes slow cooling and that the cooling
frequency is below the X-ray (if the latter is not satisfied then
cooling is not the limiting factor anyway, see below). It also
ignores the effect of IC cooling on $\nu_c$ which can only delay the
time at which KN effects becomes important. Thus, at $t\sim
10^{4}-10^{5}$s, where these conditions are typically valid,
Klein-Nishina effects are negligible and $\epsBd^{1/2} Y_x$ is of
order unity (for the same reasons discussed above equation \ref{eq
eta_IC_ISM} in the context of ISM). Thus observations of $10$ keV
photons during the first day imply:
\begin{equation}\label{eq eta_synchWind}
    \eta \lesssim 30 \left(\frac{h \nu_{obs} }{10 {~\rm keV}}\right)^{-1}
    \left(\frac{E_{53}}{A_*^3}\right)^{1/4} B_{u,\mu G,19}~ t_5^{-1/4} .
\end{equation}
This result is similar to the one obtained in ISM. It implies that
the conclusion that the observed X-ray afterglows indicate on a fast
acceleration is largely independent of the circum-burst density
profiles. The same is applicable to the conclusion that currently
there is no strong indication for amplification of the magnetic
field in the shock upstream.  Note that this limit is valid only of
the X-rays are observed to be above the cooling frequency. If X-ray
photons are not cooling over the system dynamical time (e.g., due to
a very low value of $\varepsilon_B$) then the confinement limit,
which require that the X-ray emitting electrons spend less time than
the dynamical time in the upstream, is more constraining than the IC
limit and should be used instead.

The confinement limit is unimportant at early time, but it becomes
more stringent with time and may become the dominant limit at very
late time, $t \sim 10^7$ s  or even earlier if $\epsilon_B \ll
10^{-3}$. X-ray afterglows that are observed at such late time are
very rare, but they do exist, as we discuss in the following
section.

\section{GRB 060729}\label{sec GRB 060729}
GRB 060729 is the burst (at z=0.54) with the latest X-ray detection,
642 days after the burst \citep[][ hereafter G10]{Grupe10}. The late
time X-ray emission show a temporal break, from $F_\nu \propto
t^{-(1.32^{+0.02}_{-0.05})}$ to $F_\nu \propto
t^{-(1.61^{+0.10}_{-0.06})}$, roughly $10^6$ s after the explosion.
At the same time the X-ray spectrum varies from $F_\nu \propto
\nu^{-(1.18\pm0.11)}$ to $F_\nu \propto \nu^{-(0.89\pm0.11)}$. This
simultaneous temporal break and spectral hardening fits very well
(within 1$\sigma$) a passage of the cooling frequency, which
increases with time, through the X-ray band (G10). This behavior of
increasing cooling frequency is expected in a wind external medium.
G10 find that a model of a spherical blast wave in a wind profile
medium, where $E=10^{54}$ erg (isotropic equivalent), $A_*=0.1$,
$\varepsilon_B=0.003$ and $\varepsilon_e=0.1$, fits the data well
until $t \approx 4 \cdot 10^7$s, when a very sharp temporal break is
observed. The spectral evolution during this late break is hard to
constrain, due to the faintness of the signal, but it shows
indications of softening.

The origin of the late temporal break is not well determined. G10
discuss two possible origins - a jet break or a break in the
electron distribution. They find that it is hard to reconcile the
late break with a jet origin, although they cannot rule it out. On
the other hand a spectral origin can provide a more consistent
explanation. In that case the most natural source of the temporal
break is inability of the shock to accelerate X-ray emitting
particles. In that case these observations provide the first direct
measurement (not only an upper limit) of $\eta$. Note that according
to the model of G10, the blast wave is still relativistic even a
year after the burst, $\G \approx 5$, due to the large blast wave
energy and low external density. At late time, when the cooling
frequency is above the X-ray band, the limit on acceleration of
X-ray emitting electrons must be due to confinement. Thus if indeed
the late break in the afterglow, at $t \approx 4 \cdot 10^7$s, is
due to limited acceleration then:
\begin{equation}\label{eq eta_synchWind}
    \eta \approx 100 A_{*,-1}^{1/4}{{\left( \frac{\varepsilon
_{B,d}}{0.003} \right)}^{1/4}} B_{u,\mu G,19}
\end{equation}
Hence, unless the upstream field is significantly amplified by the
shock precursor, acceleration cannot be very slow also when $\G
\approx 5$. Moreover, if $B_{u,\mu G,19} \lesssim 0.1$ then the
acceleration must be very fast and the origin of the observed break
is almost certainly due to limited acceleration. If $B_{u,\mu G,19}
\gtrsim 1$, and the break is due to limited electron acceleration,
then $\eta \gtrsim 100$ which is significantly larger than the value
suggested by earlier  X-ray ($\sim$ day) and GeV ($\sim 10^3$ s)
observations. This may suggest that the efficiency of particle
acceleration is reduced when the shock approaches mildly
relativistic velocities. Finally, even if the break is not related
at all to electron acceleration then the equality in equation
\ref{eq eta_synchWind} becomes an upper limit on $\eta$.

\section{Summary}
In this letter we examined the constraints that GRB afterglow
observations poses on the acceleration rate, within the DSA
framework, in relativistic, weakly magnetized, collisionless shocks
in ion-electron plasma. We examine shocks that propagate into a
constant density medium (ISM) and into a decreasing density of
massive stellar wind. We consider three major factors that limit the
acceleration in such shocks, confinement, synchrotron cooling and IC
cooling. We find that at early times ($\sim 10^3$ s) the best limits
are set by synchrotron cooling of Gev emitting electrons while at
intermediate times ($\sim 10^5$ s) IC cooling of X-ray emitting
electrons provides the best constraints. These results are
independent of the circum burst medium density profile. At very late
time ($\gtrsim 10^6$ s) confinement may becomes the dominant factor
in a wind environment while IC cooling remains the dominant factor
that limits the acceleration in ISM.

Examining available observations, the tightest limits are obtained
by GeV photons that are observed $100-1000$ s after the burst, if
these are synchrotron photons from the external shock. The origin of
these photons is not determined yet, but they are seen long after
the prompt emission fades and are therefore, most likely, originate
in the external shock. Various modelings of the GeV emission find
that synchrotron emission can explain the observations well
\citep{Kumar09,Kumar10,Ghisellini10}. If this is true, then the
observed GeV emission require an extremely fast acceleration, at the
Bohm limit or faster, i.e., $\eta \lesssim 1$. This limit is very
robust since it is almost independent of any of the shock parameters
such as energy, density etc. On time scales of $100-1000$ s the
shock Lorentz factor is $\sim 100$

X-ray ($\sim 10$) keV photons are regularly observed on time scales
of hours-days, where $\G \sim 10-50$. The IC cooling of these
photons also provides a tight limit on the acceleration rate: $\eta
\lesssim 15 B_{u,-5} n_0^{-5/8}$ in ISM (c.f. \citealt{Li06}) and
$\eta \lesssim 30 B_{u,\mu G,19} A_*^{-3/4}$ in a wind. On one hand
these limits are less robust than those obtained by the GeV photons,
due to the uncertainty in $B_u$ and the external density, but on the
other hand the certainty that these X-ray photons are emitted by
synchrotron process in the external blast wave is much higher. Note
that if, as suggested by the GeV data (and by the observations of
other acceleration sites such as the Crab nebula; \citealt{Abdo11}),
acceleration can be as fast as $\eta \sim 1$, then the available
X-ray observations may be consistent with no amplification of the
magnetic field in the shock upstream region (contrary to previous
conclusions of \citealt{Li06}).

Finally, GRB 060729, is the burst with the longest duration X-ray
afterglow observed to date. Its afterglow shows a sharp decline in
the integrated X-ray flux $4 \cdot 10^7$ s after the burst (G10).
This decline is most likely accompanied by a spectral softening.
This decline can be explained naturally if the synchrotron frequency
of the maximally accelerated electron is crossing the X-ray band at
$t \approx 4 \cdot 10^7$ s. If this is the case than this is a
direct measurement of $\eta$. The afterglow light curve is
consistent with a wind circum burst density and a cooling frequency
that crosses the X-ray band at $t \approx 10^6$ s, implying that at
later time the X-ray emission can be limited only by confinement.
Using the fit of G10 to the afterglow parameters, the shock Lorentz
factor at the time of the fast decline is $\approx 5$ and $\eta
\approx 100 A_{*,-1}^{1/4}{{\left( \frac{\varepsilon _{B,d}}{0.003}
\right)}^{1/4}} B_{u,\mu G,19}$. If the fast decline at $t>4 \cdot
10^7$ s is not due to shock acceleration limit than the equality
becomes an upper limit. These results suggest that the acceleration
rate remains rather fast also at lower Lorentz factors.

To conclude, we find that GeV and X-ray afterglow observations,
provide independent limits on $\eta$. The combination of these
limits strongly suggest that particle acceleration is fast in
relativistic, weakly magnetized,  collisionless shocks in
ion-electron plasma. Namely, diffusion in the shock upstream and
downstream regions take place close to the Bohm limit at $\G \approx
100$ and it remains fast during the shock deceleration, at least up
to $\G \approx 5$.

We thanks Rodolfo Barniol Duran, Zhuo Li and the anonymous referee
for helpful comments. This research was partially supported by ISF
grant No. 174/08 and by an ERC starting grant.


\end{document}